\documentclass[fleqn,usenatbib]{mnras}

\usepackage{newtxtext,newtxmath}

\usepackage[T1]{fontenc}

\DeclareRobustCommand{\VAN}[3]{#2}
\let\VANthebibliography\thebibliography
\def\thebibliography{\DeclareRobustCommand{\VAN}[3]{##3}\VANthebibliography}

\usepackage{graphicx}	
\usepackage{amsmath}	

\usepackage{graphicx} 
\usepackage{adjustbox}
\usepackage{times}
\usepackage{longtable}
\usepackage{ae,aecompl}
\graphicspath{{./Images/}} 

\newcommand{\jaavso}{Journal of the American Association of Variable Star Observers }

\title[Period changes of AHB stars]{A search for period changes of eight short-period Type II Cepheids}

\author[A. M. Yacob et al.]{
	Alemiye M. Yacob,$^{1,2,3}$\thanks{E-mail: alemiyem@essti.gov.et (AMY)}\thanks{Research visitor @ SAAO, SA}
	Leonid N. Berdnikov,$^{4}$
	Elena N. Pastukhova,$^{5}$
	Alexei Y. Kniazev,$^{3,4,6}$\newauthor
	Patricia A. Whitelock $^{3,7}$\\	
	$^{1}$ Astronomy and Astrophysics Department, Entoto Observatory and Research Center 
	(EORC),\\
	Space Science and Geospatial Institute (SSGI), P.O. Box 33679, Addis Ababa, Ethiopia\\
	$^{2}$ Addis Ababa University (AAU), P.O.Box 1176, Addis Ababa, Ethiopia\\
	$^{3}$ South African Astronomical Observatory (SAAO), P.O. Box 9, 7935 Observatory, Cape Town, South Africa\\
	$^{4}$ Sternberg Astronomical Institute (SAI) of the Moscow State University, Universitetskii pr. 13, 119992 Moscow, Russia\\
	$^{5}$Institute of Astronomy, Russian Academy of Sciences, Pyatnitskayaul. 48, 119017 Moscow, Russia\\
	$^{6}$Southern African Large Telescope (SALT) Foundation, P.O. Box 9, 7935 Observatory, Cape Town, South Africa \\
	$^{7}$ Department of Astronomy, University of Cape Town, 7701 Rondebosch, South Africa
}

\date{Accepted 2022 July 28. Received 2022 July 28; in original form 2022 June 28}

\pubyear{2022}

\begin{document}
	\label{firstpage}
	\pagerange{\pageref{firstpage}--\pageref{lastpage}}
	\maketitle
	
	\begin{abstract}
		In this study, we investigate the period changes of eight short-period Type II Cepheids of the BL Her subtype, i.e., with periods in the 1-4 day range.  The $O-C$ diagrams for these stars are constructed using all suitable  observational data from ground and space surveys. This spans a time interval of over one century and includes digitized photographic plates as well as photometry from the literature.  The $O-C$ diagrams show parabolic evolutionary trends, which indicate the presence of both increasing and decreasing periods for these eight short period stars. These period changes are in good agreement with the recent theoretical evolutionary framework and stellar evolution models for BL Her stars. The pulsation stability test proposed by Lombard and Koen also suggests that the changes in the periods are real. 
		
	\end{abstract}
	
	
	
	
	\section{Introduction}
	
	Type II Cepheids (T2Cs) are old, low-mass and radially pulsating variable stars with luminosities below the classical Cepheids and above the RR Lyraes (RRLs) in the Hertzsprung-Russell diagram  \cite[e.g.,][]{Wallerstein2002}.\\
	\par 
	T2Cs are in the post-horizontal branch evolutionary phase of low-mass stars, where double shell (hydrogen and helium) burning occurs in thin shells outside the inert carbon-oxygen core. As it ascends the asymptotic giant branch (AGB) the star expands rapidly and becomes cooler and brighter.
	It experiences shell flashes at the boundary between the core and the helium shell. During these shell flashes the radius decreases temporarily; thus, the star moves to higher temperatures and into the instability strip (IS). As these stars leave the  horizontal branch (HB) and evolve further up the AGB, they cross the IS at higher luminosity than RRLs and become T2Cs  \citep[see, e.g.,][and reference therein]{Bhardwaj2022,Catelan2015}.\\
	\par
	Based on their pulsation periods T2Cs are classified as BL Herculis (BLHs) for periods between 1 and 4 days, W Virginis (WVs) for periods greater than four and less than 20 days, and RV Tauri (RVs) stars for period greater than 20 days \cite[e.g., ][]{Soszynski2018}.
	
	\par
	The three grouping of T2Cs represent various stages of post-horizontal branch evolution. In each evolutionary
	stage, predicted stellar properties can be compared with observations. Short-period T2Cs stars crossing the IS show increasing periods as they evolve redward from the HB to the AGB. The intermediate-period WVs, evolving at higher luminosities on the AGB, suffer helium shell flashes and are
	making temporary excursions from the AGB into the IS, showing either increasing or decreasing periods. RVs stars, with long periods, are post-AGB stars undergoing thermal pulses that send them blueward
	into the IS and show only decreasing periods \citep{Gingold1976,Bono1997b,Wallerstein2002}.
	
	The period change is one of the essential properties of pulsating stars obtained directly from observational data that can be used to constrain and test stellar structure models and to probe stellar evolution. The period change is thus one of the parameters that can detect evolutionary effects on human time-scales.
	
	\par
	Several studies have been conducted to investigate the pulsation and evolutionary status of T2Cs  as described in reviews by \citet{Wallerstein2002} and \citet{Neilson2016}. A number of theoretical evolutionary track grids and pulsation models have been developed \cite[see e.g.,][]{Gingold1976,Bono1997a,Bono1997b, Bono2016, Bono2020,Smolec2016,Dotter2008,DellOmodarme2012} that can be used to compute the pulsation properties including period changes, period distributions, crossing modes and other characteristics of T2Cs in their post-HB evolution.
	\par
	From theoretical considerations we would expect short-period T2Cs (BLH) stars, to be dominated by increasing periods as the they move redward from the hot edge to the cool edge of the IS \citep[e.g.,][]{Gingold1976,Bono2016,Sandage1994,Wallerstein2002}.

	\par
	Several period change studies of BLH stars in globular clusters \citep{Osborn1969,Wehlau1982,Jurcsik2001,Osborn2019}  and in the field \citep{Christianson1983,Diethelm1996,Provencal1986} have found that the majority of BLH stars do indeed show period increases while a minority, within observational uncertainties and limited data, show no period change. However, the changes are broadly consistent with theoretically predicted period changes \citep{Neilson2016, Smith2013}. 
	
	\par 
	For BLH stars, the most apparent evolutionary trend is redward evolution with an increasing period. However, Gonzales (1994) discovered one example that demonstrated blue-ward advancement with a decreasing period. Furthermore, the current theoretical framework \citep{Bono2020} predicts that BLH
	stars can experience both negative and positive period changes during their post-HB evolution. This prediction was based on gravo-nuclear instabilities at the onset of He-shell burning, following the exhaustion of helium in the core. The existence of these \textquoteleft gravo-nuclear loops\textquoteright  (GNLs) causes BLH stars to stay longer inside the IS and make fast period changes \citep{Bono1997a,Bono1997b}. Such scenarios would suggest that BLHs could show alternately increasing and decreasing periods on quite short timescales. 
	The period changes found in our study of BLH stars support the existence of GNLS, and if interpreted in terms of breathing pulses they could provide new insight into the very uncertain  mixing processes which are crucial for stellar evolution \citep[see also][]{Sweigart2000}.
	
	\par In this work, we revise and study the period changes of eight BLH stars, with periods between 1 and 2 days, which are usually called {\it Above Horizontal Branch} (AHB) stars \citep{Diethelm1983} during their post-HB evolution. We also investigate the evolutionary changes in period of BLHs from records dating back over a century and compare them to those predicted by the theory of stellar evolution. The eight BLH stars, in order of increasing period, are V716 Oph, BF Ser, CE Her, BL Her, XX Vir, KZ Cen, V745 Oph, and V439 Oph; they are listed in Table~\ref{tab:table1}.
	
	\par This paper is structured as follows: in section~\ref{sec:two} we describe the data sources, methods and analyses used for this study. In section~\ref{sec:three} we present and discuss the result of our work by comparing it with theoretical models and previous observational analyses. Finally, section~\ref{sec:four} provides a summary and conclusion.

	\section{Data, Methods and Analysis}
	\label{sec:two}
	For this research, we used all suitable ($B$, $V$, and $g'$) band data we could find in the literature, from ground- and space-based surveys, including digitized photographic plates. Specifically, we used: (i) photographic (PG) data from  Digital Access to a Sky Century at Harvard (DASCH)~\citep{Grindlay2012}, and from the Sternberg Astronomical Institute (SAI); (ii) CCD data (filters $V$ and $g'$) from the All Sky Automated Surveys (ASAS-3)~\citep{Pojmanski2002}, the All Sky Automated Surveys of SuperNovae (ASAS-SN)~\citep{Kochanek2017}, INTEGRAL-OMC \citep{AlfonsoGarzon2012}, Catalina Sky Survey (CSS)~\citep{Drake2013}, PAN-STARSS \citep{Chambers2016}, HIPPARCOS \citep{Perryman1997}; 
	(iii) photoelectric (PHE)  (filters $B$ and $V$), visual (VIS) and photovisual (PGV) data from the literature. The photographic data, from DASCH, SAI and others, obtained from the literature are assumed to be identical to $B$-band photometry for the purpose of this analysis.
	
	We present the general information for the stars from the general
	catalogue of variable stars (GCVS) \citep{Samus2017} in Table~\ref{tab:table1}. All other data sources used for this study are listed in Table~\ref{tab:tabel2}.
	
	The data obtained from different sources are combined and sorted in order of Heliocentric Julian Date (HJD). They are then divided into seasonal batches, generally covering intervals of at least 3-months each year, although this obviously depends on the individual data sets and the length of the gaps between them. For the PG data there are sometimes gaps of several years between data-sets. The seasonal light-curves generated for each data-batch were then used to construct the $O-C$ curves.
	
	To search for any period changes we employed the observed minus calculated ($O-C$) time of maximum as illustrated in an $O-C$ diagram. The observed light-curves ($O$) were constructed for each star at each season and then compared with a reference light-curve ($C$) to calculate the difference between the times of maximum light ($O-C$). The $O-C$ residuals were determined using the \cite{Hertzsprung1919} method, with the algorithm developed by \cite{Berdnikov1992}. At the same time we correct for the shifts in the time of maxima between observations made through different filters. We use the method by \cite{Lombard1993} to confirm that the pulsation period changes are real.
	
	Assuming that the change in a period is linear with time, we can calculate the maximum  light elements for all stars using the following quadratic elements \citep{Sterken2005} : 
	
	\begin{eqnarray}\label{eq:1}
	HJD_{\textrm{max}} = M_{\textrm{o}} + PE + QE^{2} ,
	\end{eqnarray}
	where $P$ is the period at the adopted epoch, $ M_{\textrm{o}}$,  $Q$ is a parabolic term and $E$ is the number of the cycle. The quadratic coefficient $Q$ relates to $O-C$ with the following equation :
	\begin{eqnarray}\label{eq:2}
	Q = \frac{1}{2}\frac{dp}{dt}P_{\textrm{mid}}E^2, 
	\end{eqnarray}
	where $dP/dt$ is the rate of period change and $P_{\textrm{mid}}$ is the mid-epoch period.  
	
	The quadratic coefficient $Q$ can be used to measure the values of period change ($dP/dt$) in seconds per year ($s/yr$) or in days per million years ($d/Myr$) using the following relations in Eqn.~\ref{eq:3} and \ref{eq:4}, respectively, using sidereal years:
	
	\begin{eqnarray}\label{eq:3}
	\frac{dP}{dt} = 365.25\times24\times60\times60\times(\frac{2Q}{P_{\textrm{mid}}})  ,
	\end{eqnarray}
	
	\begin{eqnarray}\label{eq:4}
	\frac{dP}{dt} = 730.5\times10^{6}\times(\frac{Q}{P_{\textrm{mid}}}).
	\end{eqnarray}
	\begin{table}
		\centering
		\caption{Summary of information for the BLH stars from the GCVS.}
		\label{tab:table1}
		\begin{tabular}{|c|c|c|c|} 
			\hline
			\textbf{ Star} & \textbf{Period} & \textbf{Max $V$} & \textbf{Min $V$}\\
			& [d] & [mag] & [mag] \\
			\hline
			V716 Oph & 1.116 & 11.28 & 12.60 \\
			BF Ser   & 1.165 & 11.05 & 12.56 \\
			CE Her   & 1.209 & 11.53 & 12.92 \\
			BL Her   & 1.307 & 9.70  & 10.62 \\
			XX Vir   & 1.348 & 11.55 & 12.78 \\ 
			KZ Cen   & 1.520 & 11.80 & 12.77 \\
			V745 Oph & 1.595 & 12.70 & 13.90 \\
			V439 Oph & 1.893 & 11.73 & 12.70 \\[1ex]  
			\hline     
		\end{tabular}
	\end{table}
	\begin{table*}
		\caption{Summary of the Data Sources}
		\label{tab:tabel2}
		\begin{tabular}{lllll}
			\hline
			\textbf{Star Name} & \textbf{JD \& Year Intervals} & \textbf{No. of Observ.} & \textbf{Type of Observation}& \textbf{References}  \\
			\hline
			V716 Oph  & 2411544 - 2458589  & 2041 & PG  & DASCH                                               \\
			&     1890 -  2019          & 199                     & PG                           &   \citet{Mandel1970, Kinman1984,Kinman1965}                               \\
			&                      & 117                     & PG                           & SAI                                                                                                                                             \\
			&                      & 17                      & CCD ($g'$)                 & PAN-STARSS                                                                                                                                      \\
			&                      & 126                     & CCD ($V$)                  & INTEGRAL- OMC                                                                                                                                     \\
			&                      & 176                     & PHE ($B$,$V$)            & \citet{Kwee1984, Diethelm1986, Lin1983}                                                                                                               \\
			&                      &                         &                              & \citet{McNamara1994,Bookmeyer1977}                                                                                                         \\
			&                      & 330                     & CCD ($V$)                  & CSS                                                                                                                                              \\
			&                      & 1448                    & CCD ($g'$,$V$)           & ASAS-SN                                                                                                                                          \\
			&                      & 509                     & CCD ($V$)                  & ASAS3              2411549                                                                                                                               \\
			BF Ser             & 2411549 - 2458589    & 2226                    & PG                           & DASCH                                                                                                                                              \\
			&   1890 -  2019                    & 780                     & PG  VIS                      & \citet{Ashbrook1950,Soloviev1952,Mandel1970,Nikulina1966}                                                                               \\
			&                      & 69                      & PG                           & SAI                                                                                                                                               \\
			&                      & 11                      & CCD ($g'$)                 & PAN-STARSS                                                                                                                                       \\
			&                      & 132                     & CCD ($V$)                  & INTEGRAL-OMC                                                                                                                                      \\
			&                      & 126                     & CCD ($V$)                  & HIPPARCOS                                                                                                                                       \\
			&                      & 128                     & PHE ($B$,$V$)            & \citet{Henden1996,Diethelm1982,Harris1980}                        \\
			& 	&	 & 	&  Ashbrook (1950); Soloviev (1952); Bookmeyer et al. (1977)    
			\\	
			&                      & 367                     & CCD ($V$)                  & CSS                                                                                                                                            \\
			&                      & 1510                    & CCD ($g'$,$V$)           & ASAS-SN                                                                                                                                          \\
			&                      & 355                     & CCD ($V$)                  & ASAS3                                                                                                                                           \\
			CE Her             & 2411586 - 2458589    & 1584                    & PG                           & DASCH                                                                                                          \\
			&   1890 -  2019       & 247                     & PG         &VIS                                   \\
			&                      & 9                       & CCD ($g'$)                 & PAN-STARSS                                                                                                                                       \\
			&                      & 118                     & CCD ($V$)                  & INTEGRAL-OMC                                                                                                                                     \\
			&                      & 189                     & PHE ($B $,$V$)            & \citet{Loomis1988,Chambers2016}                                                                                 \\
			&	 & 	& 	&  Harris(1980); Diethelm \& Tammann (1982)       	  \\
			&                      & 101                     & CCD ($V$)                  & CSS                                                                                                                                              \\
			&                      & 1591                    & CCD ($g'$,$V$)           & ASAS-SN                                                                                                                                           \\
			Bl Her             & 2411508 - 2458589    & 4449    & PG           &DASCH                                   \\
			&    1890 -  2019                 & 504                     & PG \& PGV          & Mandel(1970)                                                                                                                                             \\
			
			&                      & 54                      & CCD (V)                      & INTEGRAL-OMC                                                                                                                                     \\
			&                      & 123                     & CCD ($V$)                  & HIPPARCOS                                                                                                                                         \\
			&                      & 658                     & PHE ($B$,$V$)            & \citet{Mitchell1964,Moffett1984}    \\
			& 	&	 & 	& \citet{ArellanoFerro1998,Alexander1987,Meakes1991}  	 \\
			&   &    &  & \citet{MichalowskaSmak1965,Preston1967}   \\
			& 	&	 & 	& \citet{Ignatova2000,Szabados1977}; Harris(1980)	   \\
			&                      & 1647                    & CCD ($g'$,$V$)           & ASAS-SN                                                                                                                                          \\
			&                      & 495                     & CCD ($V$)                  & ASAS3                                                                                                                                            \\
			XX Vir             & 2411529-2458589      & 1832                    & PG                           &	DASCH                                                                                                                                            \\
			&     1890 -  2019                 & 639                     & PG \& VIS                      & \citet{Oosterhoff1936}; Mandel (1970)                                                                                                                     \\
			&                      & 13                      & CCD ($g'$)                 & PAN-STARSS                                                                                                                                        \\
			&                      & 68                      & CCD ($V$)                  & INTEGRAL-OMC                                                                                                                                      \\
			&                      & 214                     & PHE ($B$,$V$)            & Bookmeyer et al. (1977), Loomis et al. (1988)                               \\
			& 	&	 & 	& Harris (1980); Mitchell et al. (1964); \citet{McNamara1994}                     	   \\
			&                      & 302                     & CCD ( $V$)                 & CSS                                                                                                                                             \\
			&                      & 1548                    & CCD ($g'$,$V$)           & ASAS-SN                                                                                                                                          \\
			KZ Cen             & 2411167 - 2458590    & 1527                    & PG                           & DASCH                                                       \\
			&   1889 -  2019                  & 190                     & PHE ($B$,$V$)            & \citet{Petersen1984,Irwin1961,Diethelm1986}                                                                       \\
			&                      & 1907                    & CCD ( $g'$,$V$ )         & ASAS-SN                                                                                                                                           \\
			&                      & 617                     & CCD($V$)                   & ASAS3                                                                                                                                              \\
			V439 Oph           &  2410929 - 2458589                    & 1626                    & PG                           & DASCH                                                                                                                                           \\
			&     1888 -  2019                   & 236                     & PGV \& VIS                      & Mandel (1970); \citet{Tsesevich1952}                                                                                                                \\
			&                      & 258                     & PG                           & SAI                                                                                                                                              \\
			&                      & 10                      & CCD ($g'$)                 & PAN-STARSS                                                                                                                                       \\
			&                      & 26                      & CCD ($V$)                  & INTEGRAL-OMC                                                                                                                                     \\
			&                      & 61                      & PHE ($B$,$V$)            & \citet{Sturch1966,Henden1980};\citet{Diethelm1982,Diethelm1986}          \\
			&                      & 1019                    & CCD($g'$,$V$)            & ASAS-SN                                                                                                                                           \\
			&                      & 177                     & CCD($V$)                   & ASAS3                                                                                                                                             \\
			V745 Oph           & 2412263 -2458589     & 1217                    & PG                           & DASCH                                                                                                                                             \\
			&   1892 -  2019                 & 193                     & PGV \& VIS                     & Mandel(1970)                                                                                                                                    \\
			&                      & 12                      & CCD($g'$)                  & PAN-STARSS                                                                                                                                     \\
			&                      & 80                      & CCD ($V$)                  & INTEGRAL-OMC                                                                                                                                    \\
			&                      & 41                      & PHE ($B$,$V$)            & \citet{Kwee1984}; Diethelm (1986)                                                                                                                        \\
			
			&                      & 114                     & CCD ($V$)                  & CSS                                                                                                                                              \\
			&                      & 1567                    & CCD ($g'$,$V$)           & ASAS-SN                                                                                                                                         \\
			\hline
		\end{tabular}
	\end{table*}	
	
	\section{Results and Discussion}
	\label{sec:three}
	
\begin{figure*}
		\includegraphics[width=2\columnwidth]{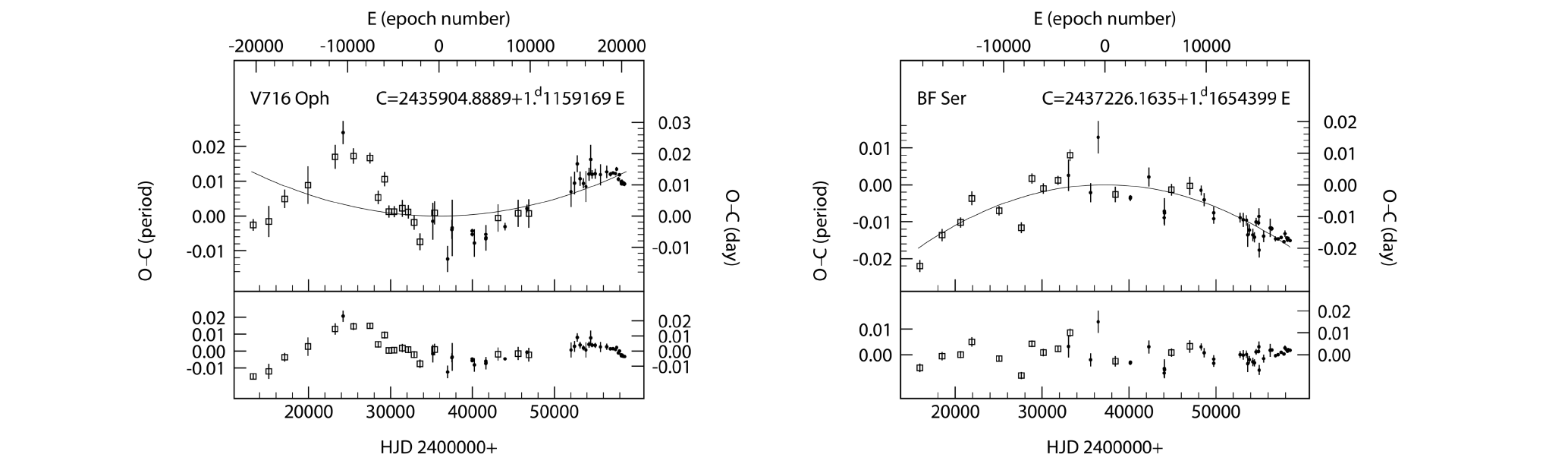}
		\vspace{0.5 cm}
		\includegraphics[width=2\columnwidth]{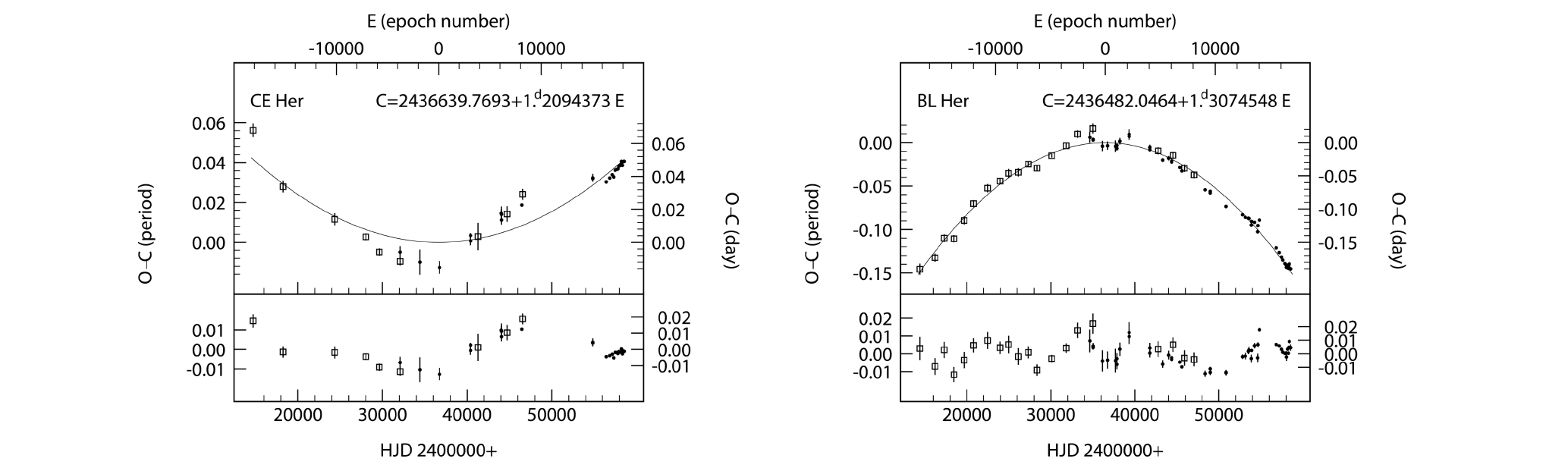}
		\vspace{0.3 cm}
		\includegraphics[width=2\columnwidth]{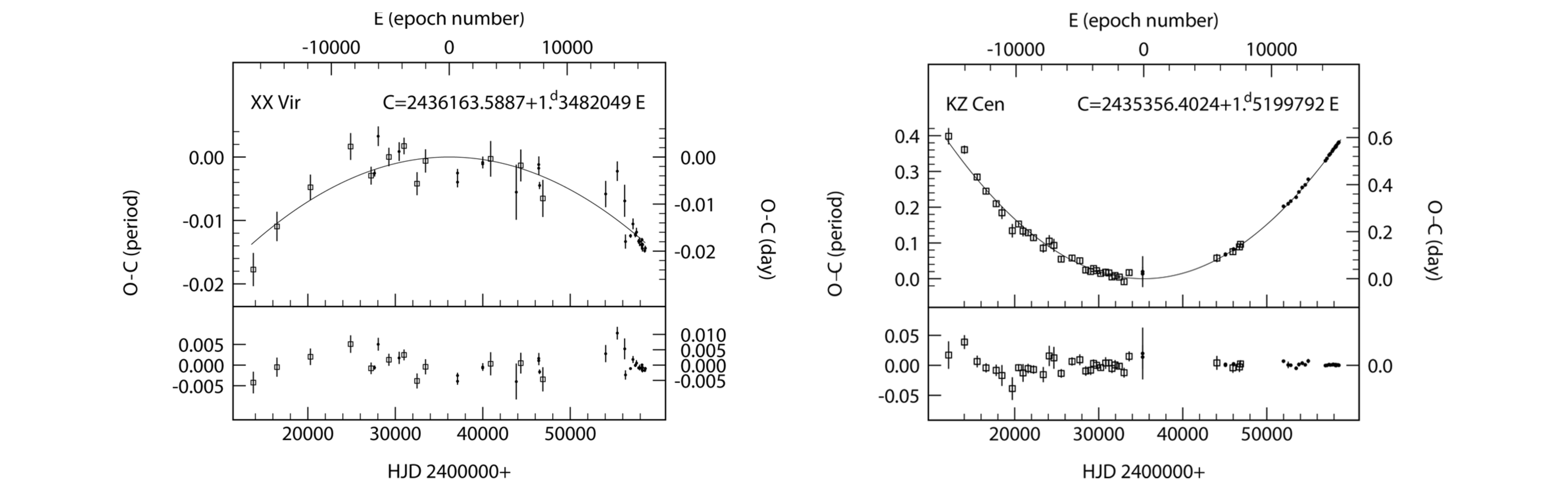}
		\vspace{0.5 cm}
		\includegraphics[width=2\columnwidth]{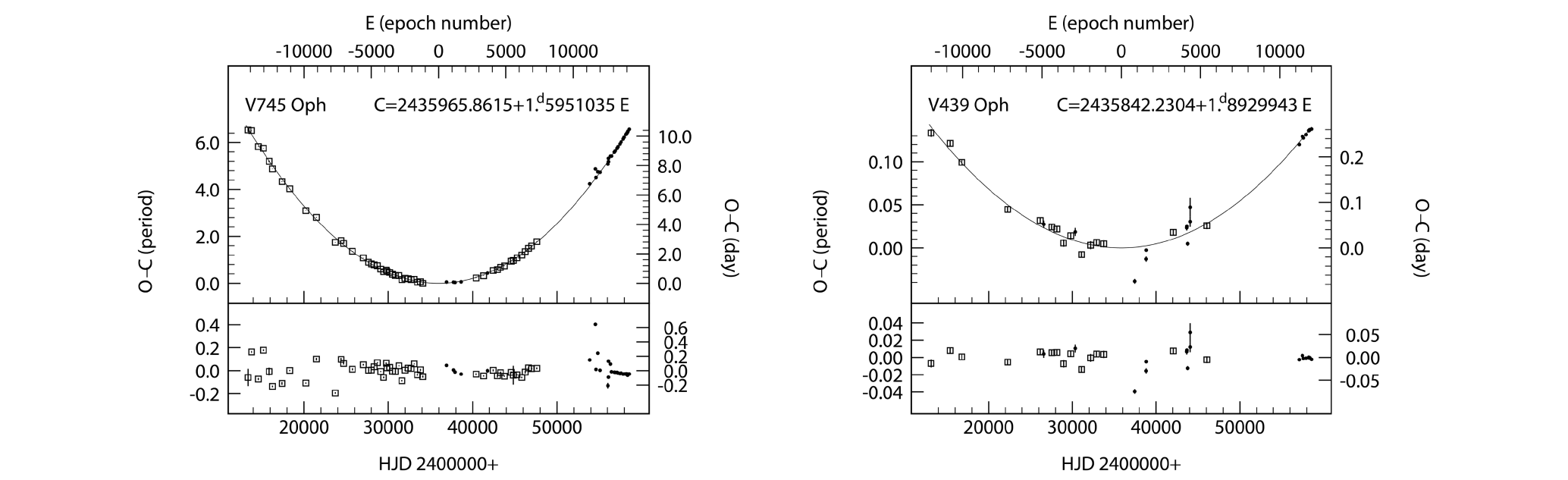}
		\caption{ The $O-C$ diagrams for eight stars relative to the linear (top) and quadratic (bottom) elements (Table~\ref{tab:table4}). The open squares represent the Harvard photographic observations, while dots are used for all other measurements; vertical bars indicating the limits of errors in the residuals.}\label{fig:1}
	\end{figure*}

\begin{figure*}
		\includegraphics[width=2\columnwidth]{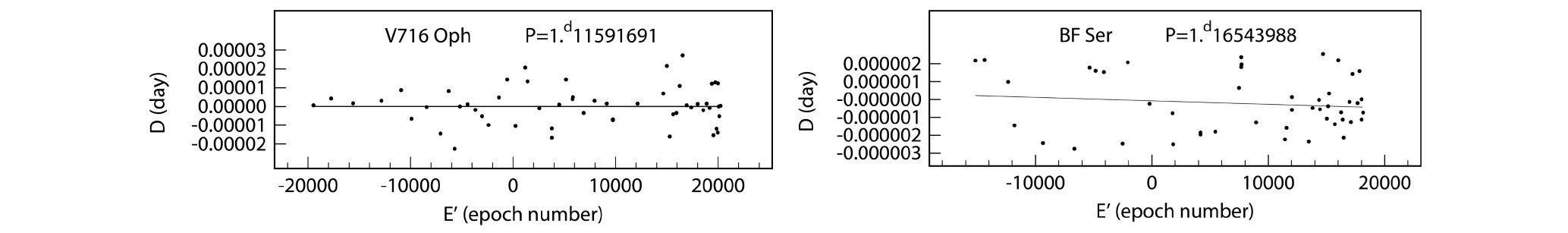}
		\vspace{0.3 cm}
		\includegraphics[width=2\columnwidth]{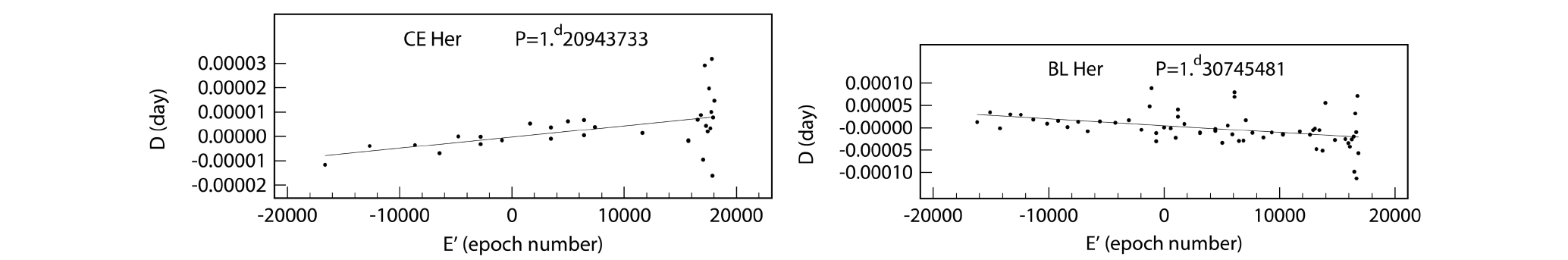}
		\vspace{0.3 cm}
		\includegraphics[width=2\columnwidth]{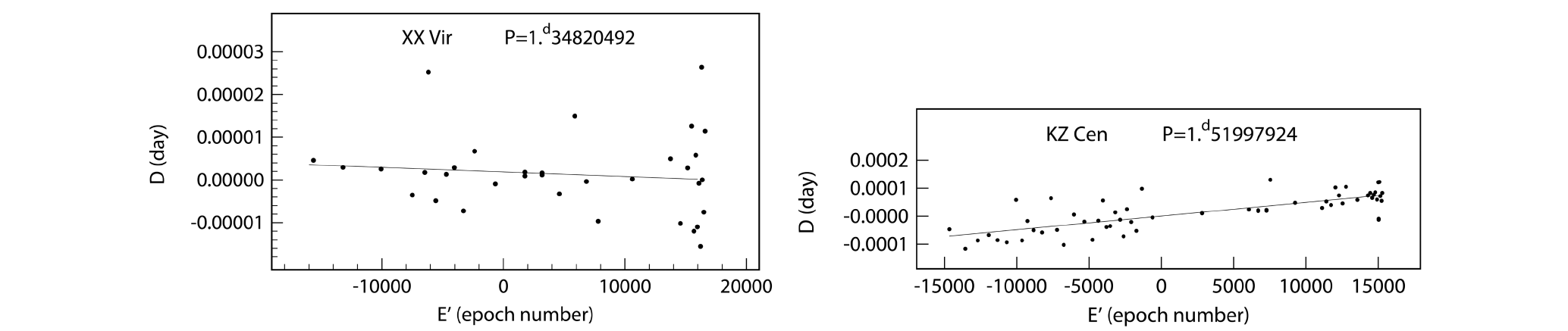}
		\vspace{0.3 cm}
		\includegraphics[width=2\columnwidth]{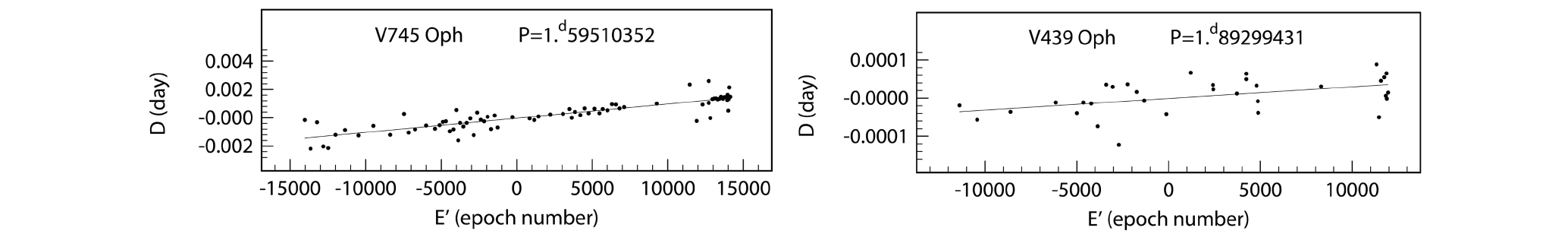}
		\caption{The period change in epoch interval ($D_{\textrm{i}}$) in days versus the epoch number ($E^{'}_{\textrm{i}}$),  showing the period change corresponding to the $O-C$ residuals. }\label{fig:2}		
	\end{figure*}

	We have calculated for the eight BLHs: the quadratic elements, the rates of period change and the corrections for maxima in the $B$- and $g'$-bands,  as listed in Tables~\ref{tab:table3} and \ref{tab:table4}. We constructed the $O-C$ diagrams for each star with all available data and performed the corresponding period change test to confirm whether the period changes are real or not. The resulting $O-C$ diagrams and the period change test graphs are presented in Figures~\ref{fig:1} and \ref{fig:2}, respectively.
	
	In the $O-C$ diagrams, we see convex parabolas (e.g., BL Her in Fig.~\ref{fig:1}) indicating increasing periods and concave parabolas (e.g., V745 Oph in Fig.~\ref{fig:1}) indicating decreasing periods. Thus we have examples of both increasing and decreasing periods within this sample of eight stars.
	
	The results from the seasonal light-curves for all our objects are presented in the appendix (supplementary materials) Table~\ref{tab:table7}. Columns 1, 2 and 3 give the times of maximum light and their errors; column 4 gives the type of observations used; columns 5 and 6 contain the epoch number E and the $O-C$ residual; columns 7 and 8 contain the number of observations N and the data sources, respectively. The data from Table A1 are shown on the $O-C$ diagrams (Fig.~\ref{fig:1}) as open squares for the Harvard photographic measurements and as dots for all the remaining observations, with vertical bars indicating the limits of the errors in the $O-C$ residuals. The parabolic elements obtained for the stars are listed in Table~\ref{tab:table3}.
	
	It is well known that the brightness maxima for pulsating variables generally occur later at longer wavelengths. This differences in the light curves from different filters was first noted by \citet{Arp1955}, who demonstrated that the times of maximum and minimum in the $V$-band ($M_{\textrm{V}}$) light-curves lag behind those in the $B$-band ($M_{\textrm{PG}}$) light-curves. When different passbands are simultaneously present in the observational data set, the primary passband should be selected in order to determine the shifts in the times of maximum light in other bands. In our case, among the three bands used ($V$, $B$, and $g'$), the $V$-band is our primary reference.
	Accordingly, we make the phase shift in the times of maximum for the $B$- and $g’$-bands with respect to the $V$-bands, which is reported in days. The phase shifts are expressed as $\Delta\phi ($B,V$)=\phi_{\textrm{$B-V$}}$ and  $\Delta\phi (\text{$g', V$})=\phi_{\textrm{$g'-V$}}$. These corrections for maxima in the $B$- and \text{$g'$}-bands are given in Table~\ref{tab:table4}. We applied them when constructing Fig.\ref{fig:1} and determining the elements (Table~\ref{tab:table4}) which refer to the $V$-band. The quadratic elements (Table~\ref{tab:table3}) make it possible to calculate the rate of change in the period ($dP/dt$), which is given in column 5 of Table~\ref{tab:table4}, together with its error.
	
	Assuming a constant rate of change for the periods We applied parabolic fits to the $O-C$ diagrams. We do see irregular changes, that do not fit the parabolic pattern, in some of the diagrams (V716 Oph, BF Ser, BL Her, Ce Her \& XX Vir); these changes could be the result of random fluctuation in the periods. Such random fluctuations have been  observed in classical Cepheid variables. The \citet{Eddington1929} method can be used to test for random, cycle-to-cycle changes in period, and it is usually applied to long-period Cepheids. For short period Cepheids, the \citet{Lombard1993}  method is useful to check whether the period change is real or due to random fluctuations. 
	
	For the BLHs we used the \citeauthor{Lombard1993} method for every star to check whether the measured $O-C$ residuals indicate real period changes. For this purpose, we calculated the differences $\Delta(O-C)_{i}$ of successive $O-C$ residuals from Table~\ref{tab:table7}, $\Delta(O-C)_{i}=(O-C)_{i+1}-(O-C)_{i}$, and plotted $D_i = \Delta(O-C)_i/(E_{i+1}-E_i)$ against $E'_i= (E_i + E_{i+1})/2$ (Fig.\ref{fig:2}). The differences, $D_{i}$, which describe the period changes in the epoch interval $E_{i}-E_{i+1}$, correspond to the behaviour of the O$-$C residuals in Fig.~\ref{fig:1}. As shown in Fig.~\ref{fig:2}, the differences $D_{i}$ in pulsation period show decreases and increases as a function of time. 
	
	From our period derivative results, we found a decreasing period for three stars, namely BF Ser, BL Her,  and XX Vir, and an increasing period for the remaining five stars, namely V716 Oph, CE Her, KZ Cen, V745 Oph, and V439 Oph. The period-change graph for V716 Oph, illustrated in Fig.~\ref{fig:2}, is almost flat, making it difficult to see if there is any significant change at all. Nevertheless, our analysis indicated an increasing period of  $0.022 \pm 0.0048$ d/Myr. In a prior study  by \citet{Diethelm1996}, XX Vir and BF Ser showed no period changes and constant period change, respectively, over 70 years. However, in the current study, we found  decreasing periods over the 129-year span. For V716 Oph and CE Her our result and \citeauthor{Diethelm1996}'s show increasing periods, although there are slightly different period changes. For the remaining stars, we found positive (KZ Cen, V439 Oph, \& V745 Oph) and negative (BL Her) period changes. The measured rate of period change in all eight BLH stars is in the range $10^{-5} $ and $ 10^{-8} $ days per year, which is the same order of magnitude as estimated by \cite{Sandage1994}, \cite{Wehlau1982}, \cite{Diethelm1996} and \cite{Osborn2019} for stars evolving from the HB towards the AGB. In addition, the parabolic trends of the $O-C$ diagrams suggests that the observed rate of period change can be understood as real evolutionary changes.

	\begin{table}
		
		\caption{The quadratic elements }
		\label{tab:table3}
		\begin{center}
			\begin{tabular}{|c|l|} 
				\hline
				\textbf{ Stars} & \textbf{$HJD_{max} = M_{\textrm{o}} + PE  +  QE^{2}$}, where $M_{\textrm{o}}$ adopted epoch, \\
				&  $P$ is the period at  adopted epoch and  $Q$ is parabolic term \\[0.5ex]
				\hline
				V716 Oph & $2435904.8889  +  1.1159169 E  +  0.34152\times 10^{-10}E^{2}$ \\
				BF Ser   & $2437226.1635  +  1.1654399 E  -  0.55228\times 10^{-10}E^{2}$ \\
				CE Her   & $2436639.7693  +  1.2094373 E  +  0.15322\times 10^{-9} E^{2}$\\
				BL Her   & $2436482.0464  +  1.3074548 E  -  0.68189\times 10^{-9} E^{2}$\\
				XX Vir   & $2436163.5888  +  1.3482049 E  -  0.68355\times 10^{-10}E^{2}$\\
				KZ Cen   & $2435356.4024  +  1.5199792 E  +  0.24842\times 10^{-8} E^{2}$\\
				V745 Oph & $2435965.8617  +  1.5951035 E  +  0.52379\times 10^{-7} E^{2}$\\
				V439 Oph & $2435842.2304  +  1.8929943 E  +  0.19312\times 10^{-8} E^{2}$\\ [1ex]    
				\hline     
			\end{tabular}
		\end{center}
		
	\end{table}

	\begin{table*}
		\begin{center}
			\caption{Corrections for maxima in the $B$- and $g'$-bands and rate of period change.}
			\label{tab:table4}
			\begin{tabular}{|c|c|c|c|c|} 
				\noalign{\smallskip}
				\hline
				\textbf{ Stars} & \textbf{Difference between $V$- \& $B$-bands} & \textbf{Difference between $V$- \& $g'$-bands} & \textbf{New Element} & \textbf{Rate of Period change}\\
				& $\Delta\phi (B,V)$[d] &$\Delta\phi (g',V)$[d] & & ($s/year$) \\[0.5ex]
				\hline
				V716 Oph &  0.0006  &  0.0048  & 2435904.8889 + 1.1159169 E &  0.00193 $\pm$ 0.00042  \\
				BF Ser   & -0.0037  &  0.0067  & 2437226.1635 + 1.1654399 E & -0.00299 $\pm$\ 0.00031 \\
				CE Her   & -0.0044  & -0.0020  & 2436639.7693 + 1.2094373 E &  0.008   $\pm$ 0.00068 \\
				BL Her   &  0.00070 & -0.0069  & 2436482.0464 + 1.3074548 E & -0.03292 $\pm$ 0.00050 \\
				XX Vir   &  0.00999 &  0.00608 & 2436163.5888 + 1.3482049 E & -0.0032  $\pm$ 0.00029 \\
				KZ Cen   & -0.0016  & -0.0018  & 2435356.4024 + 1.5199792 E &  0.10315 $\pm$ 0.00119 \\
				V745 Oph & -0.0147  & -0.0135  & 2435965.8617 + 1.5951035 E &  2.07251 $\pm$ 0.00770 \\
				V439 Oph &  0.0050  &  0.0249  & 2435842.2304 + 1.8929943 E &  0.06439 $\pm$ 0.00346  \\ [1ex]    
				\hline     
			\end{tabular}
		\end{center}
		
	\end{table*}

	\subsection{Comparison with theoretical prediction of period change by Bono et al. 2020 }
	
	Several studies have been conducted to examine the evolutionary status of post-horizontal branch stars  by constructing models and deriving evolutionarytracks for stars leaving the HB \citep[e.g.][]{Gingold1976,Bono1997a,Bono1997b,Dotter2008,DellOmodarme2012}.
	The evolutionary models of blue HB stars with depleted core helium show three different evolutionary paths depending on their mass on the HB \citep{Bono2016}. The first path is for low mass stars ($ M$/$M_{\odot} \lesssim 0.515$), these do not enter the IS and never reach the AGB. They move into their white dwarf cooling sequence and are usually referred to as AGB-Manqu\`{e} stars. The second path is for more massive stars ($0.52 \lesssim M$/$ M_{\odot} \lesssim 0.62$), these evolve redward through the IS, above the HB as they move towards their AGBs. They are known as Post-Early AGB (PEAGB). The third path is for higher mass stars ($0.62 \lesssim M$/$ M_{\odot} \lesssim 0.80$), which evolve along the AGB and into the Post-AGB phase (PAGB). They also experience thermal pulses and are therefore known as thermally pulsating AGB (TPAGB) stars. The BLH stars are thus classified as PEAGB stars that evolve redward en route to the AGB by crossing the IS \citep{Bono2020}.
	
	\par According to the evolutionary analysis of T2Cs by \cite{Gingold1976,Gingold1985}, BLHs on their way to the AGB experience blue-loops, known as \textquoteleft blue-noses \textquoteright, after crossing the IS. Before the BLHs reach the AGB the \textquoteleft blue-nose\textquoteright causes two additional crossings of the IS. Gingold argues that the reason for these three successive excursions is a readjusting of the hydrogen and helium-burning shells before the stars reached the AGB. However, recent evolutionary studies and HB evolutionary models with updated physics inputs do not support the scenario proposed by Gingold \citep{Bono2016}.
	
	\cite{Bono1997a,Bono1997b} proposed a model that involves several GNLs in the Hertzsprung-Russell diagram  that the BLH star encounters before it reaches the AGB. They pointed out that the ignition of the helium-shell in PEAGB BLH stars is dramatically different from the normal evolutionary process. Following core helium exhaustion helium-shell burning started immediately,  causing the structure to expand rapidly and creating a large discontinuity of energy sources in both the hydrogen and helium shells, that resulted in GNLs. The helium-burning shell undergoes a series of gravo-nuclear instabilities before the star resumes its normal evolutionary path to the AGB.
	
	\par \citet{Sweigart2000} investigated the GNLs reported by Bono et al. (1997a,b) during the onset of helium-shell burning at the end of the HB phase. They discovered the occurrence of GNLs, characterized by relaxation oscillations within the helium shell, that led to loops in the evolutionary tracks. \citeauthor{Sweigart2000} showed that the onset of helium-burning produces GNLs in the shell immediately following core-helium exhaustion, and indicated their strong dependence on the helium profile.
	They also argued that the treatment of convective boundaries following core helium burning (CHeB) impacts on the presence of GNLs. \citet{Constantino2016} further demonstrating that at the end of CHeB, the helium-shell encountered convective overshooting at the convective boundary, whereby the GNLs appeared during the early AGB phase. Both the \citeauthor{Sweigart2000} and \citeauthor{Constantino2016} studies supported the existence and effects of GNLs during early AGB evolution.  
	
	\par \cite{Bono2020} recently presented a theoretical framework that examines the evolutionary and pulsation properties of T2Cs based on HB evolutionary models. They predicted that BLH stars would show both positive and negative period changes. The prediction assumes the occurrence of multiple GNLs at the onset of He-shell burning. Such loops could also result in increased time spent within the IS and rapid period changes for the BLH stars in this short period regime.
	
	In our study, we found both positive and negative period changes for BLH stars during their PEAGB phases in agreement with the \citet{Bono2020} predictions.
	
	The detection of period changes that agree with the current theoretical framework proposed by Bono et al. (2020) has significance in modelling the GNLs, and many also contribute to providing detailed information about the occurrence of breathing pulses relevant to understanding Helium-burning mixing process that are important to our understanding of stellar evolution.
	
	\subsection{Comparison with predicted period change of evolutionary track model and results of Osborn et al. 2019}
	\subsubsection{Comparison with predicted period changes and crossing modes}
	
	\cite{Osborn2019} conducted a period change analysis for metal-poor BLH stars identified  in the M13 globular cluster. They calculated the observed rate of period change for three BLH stars using $O-C$ method, and found increasing  periods for all of them. 
	\par To compare the results of the measured period changes with  evolution models \citeauthor{Osborn2019} constructed colour-magnitude diagrams for M13 showing BLH and RRL stars overlaid with post-horizontal branch tracks from the models of the Pisa group \citep{DellOmodarme2012}. To understand the evolutionary stages and determine the pulsation period change rate, the Pisa evolutionary tracks are divided into three phases related to the 'blue loop'. These are: pre-loop, the first redward evolution (FRE) for the crossing of the IS, where the star's period is expected to increase; blueward evolution (BE), where the period decreases; and post-loop as the second redward evolution (SRE), where the period increases as the star resumes its evolution towards the AGB.
	
	\par
	\citeauthor{Osborn2019} estimated the rates of period change ($dP/dt$) using the Pisa models and presented a graph of $dP/dt$ versus $P$ (their fig.~9). From this diagram, all M13 stars are found to be located in the SRE phases, and the measured period changes of BLH stars are in qualitative agreement with theoretical predictions of post-HB evolution. \citeauthor{Osborn2019} argued that this $dP/dt-P$ diagram, with Pisa evolutionary tracks, can be extended and used for other BLH stars to estimate their evolutionary stage and directions while crossing the IS. Accordingly, \citeauthor{Osborn2019} determined the crossings made for 18 BLH stars with published period changes, on the $dP/dt-P$ diagram.
	\par
\begin{figure*}
		\includegraphics[width=2\columnwidth]{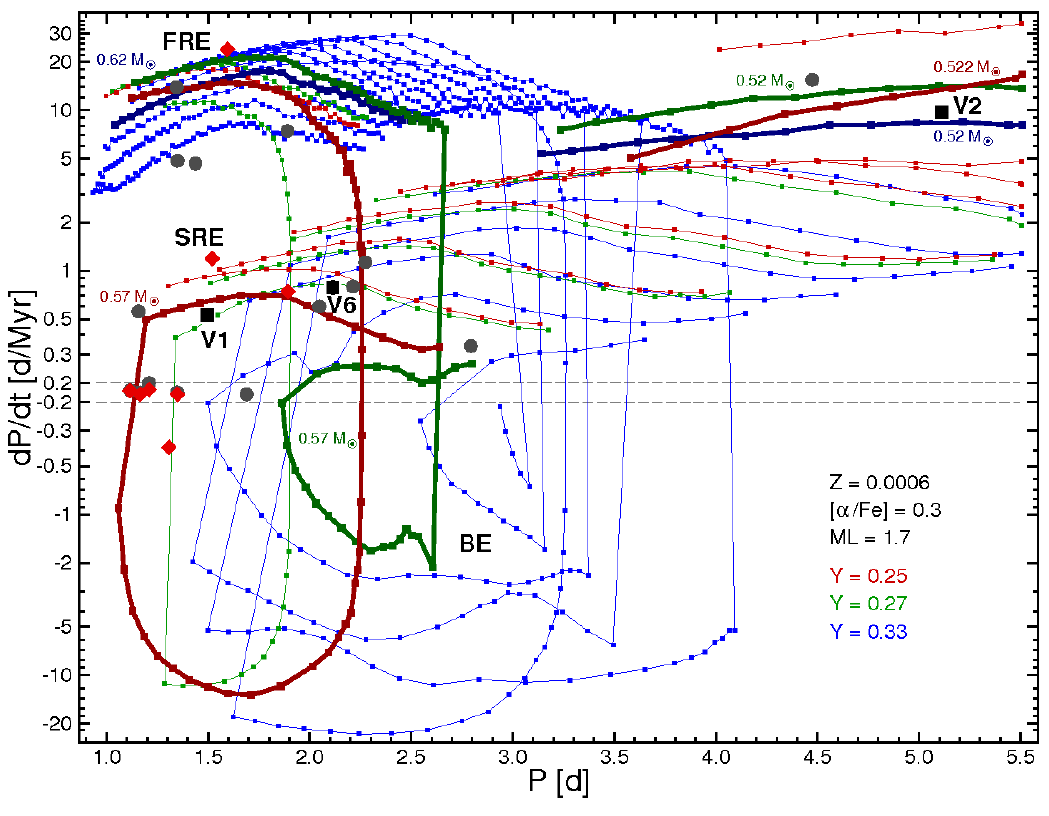}
		\caption{The theoretical rates of period change ($dP/dt$) as a function of period ($P$) for different masses ($M$) and helium abundances ($Y$), calculated from the Pisa stellar evolution models and taken from \citet{Osborn2019}. They are over-plotted with the  observed rate of period change of our BLH stars (red diamonds). Also shown are the period change rates of M13 V1, V2 and V6 (black squares), together with other T2Cs listed in Table~\ref{tab:table6} (gray circles). The ordinate shows the rate of change of the period, in days per million years, while the abscissa shows the pulsation period, in days. Models with $Y = 0.25$ are shown as red squares, with $Y = 0.27$ as green squares and with $Y = 0.33$ as blue ones. These models used $\alpha$-elements enhancement [$\alpha$/Fe]${}= +0.3$ and a mixing length parameter of ML${}= 1.7$. Models for  the same $M$ and $Y$ are connected by lines of the appropriate colour; the lines can be considered evolutionary tracks in the $P$ - $dP/dt$ plane, and the regions corresponding to the three phases of evolution produced by loops are labelled FRE, BE and SRE. Tracks with masses corresponding to the M13 Cepheids are emphasised using thick lines labelled with mass, the labels having same colour as the appropriate track. }\label{fig:3}
	\end{figure*}

	Based on the \citeauthor{Osborn2019} work, we compared our results  with the Pisa models, using evolutionary tracks with $Y = 0.25$, and indicate the crossing mode of our stars (red diamonds) in the $dP/dt-P$ diagram as shown in Fig.~ \ref{fig:3} (Osborn et al 2019, private communication 2022).

	From Fig.~\ref{fig:3}, we are able to locate our stars in the FRE, BE and SRE stages, along with the other 18 BLH stars. The crossing modes for each  evolutionary stage are presented in Table~\ref{tab:table5}; one star is in the FRE, four in the SRE and three in the BE stages. For comparison and future use we include published period change rates for BLHs with our result in Table~\ref{tab:table6}.

	
	\subsubsection{Comparison with crossing times }
	
	\par
	\citeauthor{Osborn2019} also estimated the crossing times, as a more quantitative method of comparing theory with observations, for stars evolving in the IS. Since the percentage of the crossing time spent in each evolutionary phase should correlate with the percentages of stars found in the sample, they calculated the proportion of the crossing time spent in each evolutionary phase (FRE, BE, and SRE). They compared this to the percentages of stars found in those stages in the sample, as presented in table 10 from Osborn et al. (2019). They found an absence of quantitative agreement between theory and observation, especially for BE stars. They speculated that this was due to their small sample size and proposed an in-depth examination of a larger sample of stars.
	
	\par Although we have only 3 BE stars, that is more than \citeauthor{Osborn2019} and our sample is large enough to compare crossing times quantitatively. Taking the 22 BLH stars from Table~\ref{tab:table6}, where 14 are in SRE, 5 in FRE and 3 in BE crossing modes, the percentage of stars in each phase is 64\% SRE, 23\% FRE and 14\% BE. These numbers can be compared with the theoretical values of 78\%, 13\% and 9\% (Osborn et al. (2019) table 10, for the canonical Y=0.25). This is reasonable agreement given the sample size. Obviously a larger sample would allow a more precise comparison.

	\begin{table}
		\caption{The evolutionary crossing modes of the eight BLH stars based on the $dP/dt-P$ diagram of \citet{Osborn2019} illustrated in Fig~\ref{fig:3} }
		
		\label{tab:table5}
		\begin{tabular}{|l|l|l|l|} 
			\hline
			\textbf{ Stars} & \textbf{Period} & \textbf{Rate of period change} & \textbf{Crossing modes} \\[0.5ex]
			& [d] & $dP/dt$[d/Myr] & \\
			\hline
			V716 Oph & 1.1159169 & 0.022 $\pm$ 0.0048 & SRE\\
			BF Ser & 1.1654399 & -0.035 $\pm$ 0.0035 & BE\\
			CE Her & 1.2094373 &  0.092 $\pm$ 0.0078 & SRE\\
			BL Her & 1.3074548 & -0.381 $\pm$ 0.0057 & BE\\
			XX Vir & 1.3482049 & -0.037 $\pm$ 0.0033 & BE \\ 
			KZ Cen & 1.5199792 &  1.193 $\pm$ 0.0137 & SRE\\
			V745 Oph & 1.5951035 & 23.987 $\pm$ 0.0891 & FRE\\
			V439 OPH & 1.8929943 &  0.745 $\pm$ 0.0400 & SRE\\[1ex]  
			\hline     
		\end{tabular}
	\end{table}
	
	
	\begin{table*}
		\centering
		\caption{Published period changes of BLH stars, including the current work shown in Fig.\ref{fig:3}, for comparison with Osborn et al. (2019)}
		\label{tab:table6}
		\begin{tabular}{|l|l|c|l|c|c|l|} 
			\hline
			\textbf{ Star} & \textbf{Period}($P$) & \textbf{Rate of period change}($dP/dt$) & \textbf{Error }& \textbf{Years of span} ($ \vartriangle T$) & \textbf{Crossing modes }  & \textbf{Source} \\
			& [d] & [d/Myr] & [d/Myr] & [yr]&  & \\[0.5ex]
			\hline
			\textcolor{blue}{V716 Oph} & 1.116  & 0.03[1];\textcolor{blue}{0.022[2] }& \textcolor{blue}{0.0048 [2]} &   76 [1] ;\textcolor{blue}{129[2]}& SRE & $^{1}$ Diethelm (1996) ;$^{2}$\textcolor{blue}{This paper}\\
			$\omega$ Cen V43   & 1.157 & 0.56 & 0.14 & 79 & SRE & Jurcsik et al. (2001) \\
			\textcolor{blue}{BF Ser} &1.165 & 0.0[1];\textcolor{blue}{$-$0.035[2]} & \textcolor{blue}{0.0035} & 60[1];\textcolor{blue}{129[2] } & BE & $^{1}$ Diethelm (1996); $^{2}$\textcolor{blue}{This paper} \\
			
			\textcolor{blue}{CE Her} & 1.209 & 0.2[1] ;\textcolor{blue}{0.092[2]} & \textcolor{blue}{0.0078[2]} & 65[1];\textcolor{blue}{129[2] } & SRE  & $^{1}$ Diethelm (1996) ; $^{2}$\textcolor{blue}{This paper}\\
			\textcolor{blue}{BL Her }& 1.307 & \textcolor{blue}{$-$0.381} & \textcolor{blue}{0.0057}  & \textcolor{blue}{129} & BE & \textcolor{blue}{This paper}\\
			$\omega$ Cen V92  &1.345&13.94&0.56&100 & FRE & Jurcsik et al. (2001)  \\
			\textcolor{blue}{XX Vir} &1.348 & 0.0[1];\textcolor{blue}{$-$0.037[2]} & \textcolor{blue}{0.0033[2]}&72[1] ;\textcolor{blue}{129[2]}& BE & $^{1}$ Diethelm (1996);$^{2}$\textcolor{blue}{This paper} \\
			$\omega$ Cen V60  &1.349& 4.85&0.92&100 & FRE & Jurcsik et al. (2001)  \\
			M15 V1 &1.438& 4.67&0.23&72 & FRE & Wehlau \& Bohlender (1982)  \\
			M13 V1 & 1.459 & 0.52& 0.09 & 114 & SRE & Osborn et al. (2019) \\
			\textcolor{blue}{KZ Cen} & 1.519 & \textcolor{blue}{1.193} & \textcolor{blue}{0.0137 }& \textcolor{blue}{130} & SRE & \textcolor{blue}{This paper}\\
			\textcolor{blue}{	V745 Oph}& 1.595 & \textcolor{blue}{23.987} & \textcolor{blue}{0.0891} & \textcolor{blue}{127}& FRE & \textcolor{blue}{This paper}\\
			M22 V11 & 1.690 & 0.01& 0.19 & 83 & SRE & Wehlau \& Bohlender (1982) \\
			M14 V76 & 1.890 & 7.43 & 1.0 & 48 & FRE & Wehlau \& Froelich (1994)  \\
			\textcolor{blue}{V439 Oph} & 1.893 &\textcolor{blue}{0.745}  &\textcolor{blue}{0.0400 } & \textcolor{blue}{131} & SRE & \textcolor{blue}{This paper}\\
			EK Del &2.047& 0.6&--&61 & SRE & \cite{Diethelm1990,Diethelm1996}\\
			M13 V6 & 2.113 & 0.80 & 0.06 & 114 & SRE & Osborn et al. (2019)  \\
			UY Eri & 2.213&0.8&--&66 & SRE & Diethelm (1996) \\
			$\omega$ Cen V61 &2.274& 1.13 & 0.16 & 100 & SRE & Jurcsik et al. (2001)   \\
			M14 V2 &2.794 & 0.34& 0.3 &48 & SRE & \cite{Wehlau1994}   \\
			$\omega$ Cen V48 &4.474&15.45&--&79 & SRE & Jurcsik et al. (2001)  \\
			M13 V2 & 5.111 & 9.2 & 0.7 & 115 & SRE & Osborn et al. (2019)  \\	
			\hline     
		\end{tabular}
	\end{table*}		

	\section{Summary and Conclusions}
	\label{sec:four}
	We investigate the period changes of eight BLH stars with periods between 1 and 2 days using all previously published data and all relevant observational survey data, including digitized photographic plates.
	
	The $O-C$ diagrams for the stars studied show parabolic evolutionary trends, which indicate the presence of both increasing and decreasing  periods for these BLHs over a baseline of more than a century. The derived period changes for such stars are in good agreement with the recent theoretical evolutionary framework proposed by \cite{Bono2020}, and with  the Pisa group evolutionary tracks from \citet{DellOmodarme2012} as presented by \citet{Osborn2019}.
	
	Our study found three BLH stars with decreasing periods, a significant increase over those reported by \cite{Gonzalez1994}. A more extensive study (in preparation) will further increase the number of BLH stars and improve the sample size for future studies. 
	
	We were able to determine the crossing mode of these stars by comparing with evolutionary tracks, as illustrated by Osborn et al. (2019). We found one star in FRE, four stars in SRE and three stars in BE crossing modes, respectively. The existence of decreasing periods for BLHs is important for the following reasons:
	\begin{itemize}
		\item It provides observational evidence for BE crossings with decreasing periods for BLHs, which were previously only known to have increasing periods and to be evolving redward.
		\item It fits  with the recent theoretical evolutionary framework  proposed by \citet{Bono2020}.
		\item It supports the presence of GNLs predicted by \citet{Bono1997a,Bono1997b}  and provides observational evidence for breathing pulse as suggested by \citet{Sweigart2000}. 
		\item The availability of a sample of BE stars provides quantitative agreement between theory and observation, especially for BE stars, thus the proportion of crossing times spent and the number of stars available in each evolutionary phase across the IS are approximately matched.
	\end{itemize}
	
	\section*{Acknowledgements}
	
	We thank the anonymous referee for valuable comments that improved the paper significantly. AMY acknowledges and thanks the South African Astronomical Observatory (SAAO) in Capetown, South Africa, and the Sternberg Astronomical Institute (SAI) of Moscow State University in Russia for the continuous support provided by research administrators and staff during his research visit. AMY thanks and is indebted to Space Science and Geospatial Institute (SSGI) in Ethiopia for all financial and logistical support. AMY acknowledges and thanks Prof. Wayne Osborn from Central Michigan University, and Prof. Grzegorz Kopacki, from Wroclaw University Astronomical Institute, for providing the $dP/dt$-$P$ diagram \& information calculated from Pisa, stellar evolution model. This work makes use of the observational data available from the catalogue of Digital Access to a Sky Century at Harvard (DASCH), All Sky Automated Surveys (ASAS-3), All Sky Automated Surveys of SuperNovae (ASAS-SN), INTEGRAL-OMC, Catalina Sky Survey(CSS), PAN-STARSS, HIPPARCOS and data from  Sternberg Astronomical Institute (SAI).
	\section*{Data Availability}
	
	The data underlying this article are available in the article as well as the data source cited in the article and references therein.

	
	
	\bibliographystyle{mnras}
	\bibliography{A_M_Yacob_main} 

	
	
	
	\appendix
	
	\section{supplementary materials}
	\subsection{Times of maximum light}
	
	The results of reduction of the seasonal light-curves for all stars are presented in Table ~\ref{tab:table7}. Columns 1, 2 and 3 give the times of maximum light and their errors; column 4 gives the type of observations used; columns 5 and 6 contain the epoch number (E) and the $O-C$ residual; columns 7 and 8 contain the number of observations (N) and the
	data sources, respectively.
	\onecolumn
	\begin{center}  
\begin{longtable}{c|c|c|c|c|c|l|l}                       
			\caption{Times of maximum light }\label{tab:table7}\\
			
			\hline 
			\textbf{Max, HJD} & \textbf{Year} & \textbf{Uncertainty} & \textbf{Filter} & \textbf{E } & \textbf{O\hspace{0.1mm}--\hspace{0.7mm}C}, days &  
			\textbf{N} & \textbf{ Reference }\\
			
			\noalign{\smallskip} \hline \noalign{\medskip}                                                                                                                                                                                                                                                               
			\endfirsthead
			\multicolumn{8}{c}%
			{\tablename\ \thetable\ -- \textit{Continued}} \\
			\hline
			\textbf{Max, HJD} & \textbf{Year} & \textbf{Uncertainty} & \textbf{Filter} & \textbf{E } & \textbf{O\hspace{0.1mm}--\hspace{0.7mm}C}, days &  
			\textbf{N} & \textbf{ Reference }\\
			\noalign{\smallskip} \hline \noalign{\medskip}                                                                                                                                                                                                                                                              
			\endhead
			\hline 
			\endfoot
			\hline
			\endlastfoot

			\noalign{\smallskip}
			\multicolumn{7}{ c } {  V716 Oph              }\\
			\noalign{\smallskip}
			
			2413238.3811 & 1895-02-13 &  0.0018 & $     PG $ & $  -20312 $ & $    -0.0028 $ &  31 & DASCH \\                                                                                         
			2415142.1364 & 1900-05-02 &  0.0050 & $     PG $ & $  -18606 $ & $    -0.0017 $ &  69 & DASCH \\                                                                                         
			2417078.2595 & 1905-08-20 &  0.0030 & $     PG $ & $  -16871 $ & $     0.0055 $ & 106 & DASCH \\                                                                                         
			2419917.1566 & 1913-05-29 &  0.0061 & $     PG $ & $  -14327 $ & $     0.0099 $ &  92 & DASCH \\                                                                                         
			2423221.3956 & 1922-06-15 &  0.0038 & $     PG $ & $  -11366 $ & $     0.0190 $ &  95 & DASCH \\                                                                                         
			2424211.2217 & 1925-03-01 &  0.0037 & $     PG $ & $  -10479 $ & $     0.0268 $ &  45 & \citet{Lin1983} \\                                                                                    
			2425483.3594 & 1928-08-24 &  0.0025 & $     PG $ & $   -9339 $ & $     0.0193 $ & 122 & DASCH \\
\hline
		\end{longtable}
	\end{center}
	
	\bsp	
	\label{lastpage}
\end{document}